\documentclass[12pt]{article}
\usepackage{times}
\usepackage{geometry}
\usepackage[utf8]{inputenc}
\usepackage{enumitem,amssymb}
\usepackage{ragged2e}
\usepackage{indentfirst}
\usepackage{mdframed}
\usepackage{boxedminipage}
\usepackage{multicol}
\newlist{thematic}{itemize}{8}
\setlist[thematic]{label=$\square$}
\usepackage{pifont}
\usepackage[hidelinks]{hyperref}

\usepackage{titlesec}
\usepackage[final]{pdfpages}
\usepackage{background}

\titleformat*{\section}{\Large\bfseries}
\titleformat*{\subsection}{\large\bfseries}
\titleformat*{\subsubsection}{\normalsize\bfseries}

\backgroundsetup{
scale=1,
angle=0,
opacity=1,
firstpage=false,
contents={\ifnum\value{page}=1 \includegraphics[width=\paperwidth,height=\paperheight]{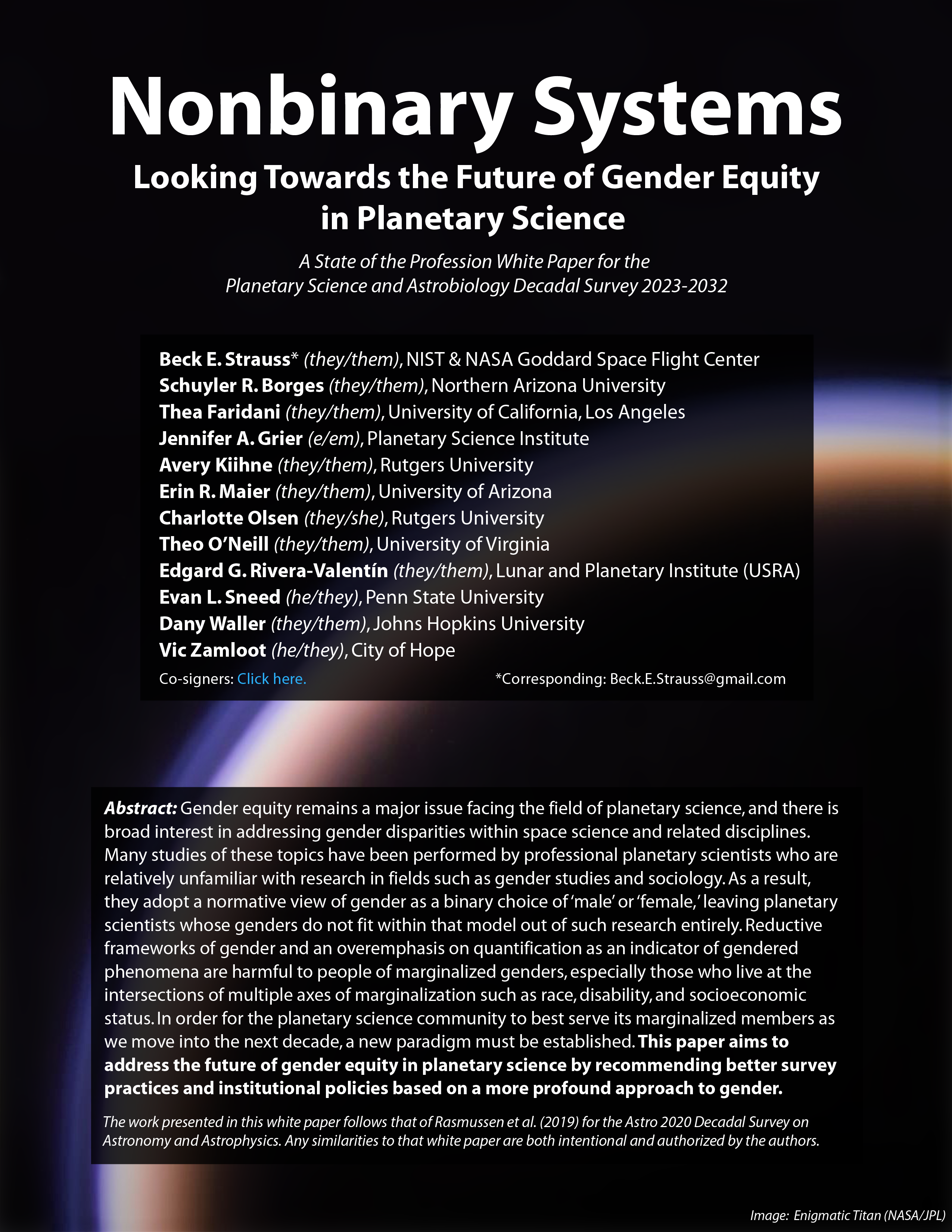}\else\fi}
}

\begin{document}

\hspace{1em}\newpage
\raggedright

\section{Introduction}\vspace{-.4em}

Over the past several decades, there have been numerous studies, community efforts, and institutional initiatives addressing the status of women\footnote{In practice, the primary subjects and beneficiaries of such work have typically been limited to cisgender, white, heterosexual, abled women.} within planetary science and astrobiology. Recently, nonbinary and transgender identities have also gained recognition and inclusion in space science equity efforts. For instance, following the American Physical Society (APS) report ``LGBT Climate in Physics: Building an Inclusive Community" [1], the American Astronomical Society (AAS) Committee for Sexual Orientation and Gender Minorities in Astronomy (SGMA) in collaboration with LGBT+ Physicists published ``LGBT+ Inclusivity in Physics and Astronomy: A Best Practices Guide" [2]. Several conferences have been organized with a focus on planetary scientists of marginalized genders, including the Women In Space Conference and the Women in Planetary Science and Exploration conference. Other major planetary science conferences with have begun to offer the option of displaying pronouns\footnote{Here, we use the term, ``pronouns" to mean the set of pronouns that should be used to refer to an individual in the third person, such as ``they/them/theirs" or ``she/her/hers". See \url{https://www.mypronouns.org/}.} on attendee badges. Networks such as 500 Queer Scientists\footnote{See the Earth and Planetary Science category at \url{https://500queerscientists.com/}.} have further increased the visibility of earth and planetary scientists of marginalized genders.\vspace{.4em}

However, many gender-related equity initiatives and studies have been led by professional planetary scientists with little to no background or training in sociology or gender studies. As a result, these studies often use methodologies that range from overly simplistic to actively harmful. In this white paper, we summarize recent work on gender equity in the field of planetary science. We then discuss common pitfalls of such work, with a focus on its detrimental effects on the inclusion of nonbinary people, replicating the findings of the Astro 2020 Decadal Survey white paper entitled ``The Nonbinary Fraction: Looking Towards the Future of Gender Equity in Astronomy'' [3]. Finally, we offer recommendations for studying gender dynamics and promoting gender inclusion in planetary science going forward into the next decade.\footnote{For a more general consideration of the necessity of and avenues for collaboration between the space and social sciences, see [4].}\vspace{.4em}

Throughout this work, we use `nonbinary' as an umbrella term for all genders not represented by the categories of `male' or `female,' following [3]. While most of our author list identifies in this way, we would like to make explicit that not everyone whose gender falls under this broad definition uses the term `nonbinary' to describe themselves, and that language surrounding gender identity is continually evolving and rarely universally agreed upon by those it purports to describe.\vspace{.4em}

\textit{N.B.: The authors wish to clarify that although we are nonbinary, we are in no way representative of all nonbinary people and should not be misconstrued as speaking for all people who experience gendered marginalization. We also acknowledge that while we emphasize the importance of the expertise that the social sciences bring to bear on this topic, we ourselves have collectively little formal training in such disciplines.}

\section{The state of gender-related studies in astronomy}\vspace{-.4em}

Several recent studies have attempted to evaluate and address gender disparities within planetary science and space science more broadly. Most of the resulting publications were written by professional planetary scientists and astronomers, and all were intended for audiences composed of space science researchers. While the literature review presented here is far from comprehensive, it is broadly illustrative of the concepts and methods that planetary scientists employ in studying gender-related phenomena within the field. (For an overview of other recent studies of gender equity in space science, we again refer the reader to [3].)\vspace{.4em}

Many of these studies examine the impact of gender on career-related metrics. For example, [5, 6, 7, 8, 9] analyze the diversity of NASA-funded spacecraft science mission teams. [10] quantifies gender discrepancies in first authorship. Another topic of focus is harassment in professional settings. The essential work of [11] examines gender and race in experiences of harassment in planetary science and astronomy, while [12] expands upon that work to consider sexual orientation in addition to gender and race. [13] reports on sexual harassment among attendees of the APS Conference for Undergraduate Women in Physics.\vspace{-.9em}

\subsection{Problematic approaches to gender and their impacts}

[3] identified three major concerns common to analyses presented in both astronomy-related papers and the planetary science-focused endeavors described here: ``(1) the treatment of gender as observable through means other than self-identification; (2) categorization schemes with limited gender options; and most critically, (3) an over-reliance on quantitative methodology that is at best insufficient for understanding gendered phenomena in [space science] and at worst epistemically violent towards people whose genders are poorly represented by these schema.'' We add a fourth concern prevalent in the planetary science literature: (4) the treatment of nonbinary and transgender identities as inconsequential and therefore dismissable.\vspace{.4em}

Many of these studies demonstrate little, if any, engagement with the enormous bodies of existing work in relevant disciplines like gender studies, transgender studies, and sociology. They further fail to prioritize the participatory inclusion of marginalized people. These methodological choices negatively impact not only nonbinary people, but people of all marginalized genders.\vspace{-.75em}

\subsubsection{Gender as observable}

Almost all of these studies, either explicitly or implicitly, rely on gender information acquired by means other than participant self-identification. Most often, subjects are assigned a binary gender using first names; in some cases, this is based upon the authors’ own perceptions, while others make use of automated methods including (but not limited to) data scraping and algorithmic analysis. In cases where gender is indeterminate based on name alone, some studies search for other public records, such as photos or articles including third-person gendered pronouns, in order to infer gender. Most simply remove data points with indeterminate gender identities.\footnote{Note that some automated systems use U.S.- or Europe-based name databases and discard all `anomalies.'}\vspace{.4em}

This treatment of gender as trivially discernible through names, physiology, or other external characteristics is inherently discriminatory.\footnote{For extensive discussion of flaws in the premises of algorithmic gender determination and the negative consequences transgender people experience from it, see [14]} For nonbinary people in particular, there is no acceptable outcome: we are either misclassified into a binary gender, or considered uncategorizable and discarded. Although this result may sound trivial to the cisgender reader, experiences of misgendering and erasure have real psychological and professional consequences for transgender, nonbinary, and gender non-conforming individuals [15, 16, 17, 18, 19, 20, 21].\footnote{For example, the authors of this paper have been so aggrieved by the studies described here that we collectively assembled, organized, and wrote not one but \textit{two} decadal white papers about it.}\vspace{-.75em}

\subsubsection{Gender as discrete}

These works tend to treat gender as a set of discrete categories that are implicitly presumed to be stable and coherent across populations, within individuals, and over time. The majority employ the male/female binary as a matter of course. Some do not require gender to be strictly binary but still require that it be discret\emph{izable}, e.g., through the selection of one of a limited number of options in surveys. Such frameworks reduce members of a category to interchangeable data points, which loses nuance and ultimately denies subjects authority over how they are represented.\vspace{-.75em}

\subsubsection{Gender as statistic}

Several of these works include statements to the effect of: ``While we recognize that gender is not binary, we do not include nonbinary people in our analysis due to lack of statistical significance.''\footnote{We recognize that such disclaimers are well-intentioned. However, we do not consider them in our assessments of said papers due to their lack of significance as anything other than empty gestures.} Statistical significance is here the determining factor in who gets to be accounted for---who \emph{counts}. Reducing the work of inclusion to that which is quantifiable and measurable produces simplistic results that fail to describe the deep nuance and complexity of gender and the experiences of people navigating it within planetary science.\vspace{.4em}

Additionally, such complexities cannot be properly understood without also considering race, dis/ability, and other axes of marginalization [22, 23, 24, 25, 26]. Even the term ``gender" has several shades of meaning: it can refer to the way someone is perceived, the way they are treated, and/or the way they see themselves [27] These concepts are all heavily influenced by other aspects of identity and social context, and are far from invariant across cultures [28, 29, 30, 31, 32, 33]. Conversations about gender equity must reckon with gender \emph{complexity}.\vspace{.4em}

Furthermore, these disclaimer statements are, in fact, \emph{incorrect}. There are dozens of nonbinary people conducting active research in space science. Readers should note the author lists of both this white paper and [3], and consult the survey results summarized in Section 3.2.1.: \textbf{between 0.5$\%$ and 13$\%$ of space scientists self-identify as nonbinary}. To dismiss these scientists as statistically insignificant is to be wrong. However, we reiterate that statistical significance must not be the first \emph{or} last word in discussions about equity and inclusion, and to give it such weight is to perpetuate harm against the most marginalized people in our communities.\vspace{-.75em}

\subsubsection{Gender as inconsequential}

Many of the efforts described here are named for and focus on the experiences of women.\textsuperscript{1} Several expand their definition of scope to groupings like ``women and nonbinary people'' or ``women+,'' terms that inadvertently situate nonbinary people as a subcategory of women and dismiss them as `basically women' or `women lite.' Related phrases like ``female and female-identifying people'' compound this effect by attempting to separate transgender women from cisgender women and thereby incorrectly implying that they are not women. \textit{(N.B.: The authors remind the reader that transgender women are women, and nonbinary people are not women, except for those who personally assert that they are nonbinary women.)} Even when such ostensibly `inclusive' language is used, these spaces tend to center cisgender women while ostracising, and ultimately reinforcing the marginalization of, the people that they purport to include.\footnote{Further discussion and recommendations for organizers may be found at \url{https://medium.com/@quinncrossley/uplifting-diverse-genders-beyond-women-and-non-binary-916c890f2185}.}\vspace{-.75em}

\subsection{Reinventing the wheel}

In the words of [3], ``Gender is not a new area of study and gender inclusion is not a new problem.'' The field of gender studies has existed for decades, and disciplines related to planetary science and astrobiology\footnote{See [34] for a particularly excellent reevaluation of approaches to sex and sexual behavior in biology.} have been working to improve gender equity for years. However, rather than deferring to established and well-researched understandings of gender and its individual and structural manifestations, many planetary scientists have chosen to approach problems of gender inclusion without consulting either this work or the experts who conduct it. As a result, they repeat lessons and rediscover concepts that are commonplace in other fields and often cause harm to those they intended to help. Furthermore, this approach incorrectly frames the problem as one whose solution does not require the expertise of marginalized people themselves.\vspace{.4em}

Since the release of [3], multiple studies of demographic diversity have been published that acknowledge the existence of that paper, note the recommendations therein, and then follow none of those recommendations.\footnote{We acknowledge the existence of such papers, but do not consider them to indicate progress on this topic and therefore do not recognize them here.}\vspace{-.9em}

\section{Recommendations}\vspace{-.4em}

Despite the flawed approaches described here, we recognize that the studies and other efforts that we have discussed come from a genuine desire to bring about positive changes within our fields. Therefore, we make the following recommendations with the aim of advancing gender equity in planetary science and astrobiology in the next decade and beyond.\vspace{-.75em}

\subsection{Methodological choices}

Approaches to gender, whether quantitative or qualitative, cannot come without a deep awareness of the complexity and context-dependence of gender. Prior to any gender-related research, responsible researchers \emph{must} begin by thoroughly grounding the precise definition of `gender' they seek to employ or investigate, the alignment of their data collection and analysis methods with said meaning, and the ways and contexts in which they intend to use the data, as well as the ways and contexts in which the data is, and is not, intended to be used.\vspace{.4em}

We do not propose to do away with quantitative methods completely; rather, we call for a community-wide reconsideration of the epistemic authority of such methods in matters of marginalization. Qualitative data and methods such as ethnographic description and participant testimony must be understood as valuable and funded accordingly [11, 35, 36, 37, 38].\vspace{-.75em}

\subsection{Collecting and reporting gender data}

\textbf{We emphatically repeat the strong recommendation of [3] discouraging the gathering of gender data through any means other than voluntary self-identification, and especially discourage the use of automated gender classification methods.}\footnote{In cases where automated gender classification is unavoidable, we recommend the R package Gendrendr, available at \url{https://dapperstats.github.io/gendrendr/.}} We encourage the Decadal Survey Committee to recommend that journals and funding agencies prioritize gender equity studies and initiatives that use the best practices described in this paper and/or are conducted in collaboration with social scientists.\footnote{See also [39] calling for improvements to data collection in earth science.}\vspace{.4em}

In all cases we strongly recommend the funded support of trained social scientists when studying marginalized people in planetary science. Therefore, \textbf{NASA should create and support dedicated funding sources for interdisciplinary research on the planetary science and astrobiology workforce, with a focus on enabling collaboration between space and social scientists by providing funding to experts in these fields}.\vspace{.4em}

For the collection of demographic data, we recommend the use of the template provided by the Open Demographics initiative.\footnote{\url{http://nikkistevens.com/open-demographics/questions/gender.html}} For further reading on gender-inclusive data collection and survey design, we direct the reader to [3] and references therein.\vspace{-.75em}

\subsubsection{Impacts of improved methods}

Several recent surveys of planetary scientists were designed to include nonbinary genders and related identities and have reported these results in a manner consistent with [3].\footnote{We also commend the work of K. Acosta on the Seminar Diversity Initiative (\url{https://diversity.ldeo.columbia.edu/seminardiversity}) at Lamont-Doherty Earth Observatory, which has aimed to improve the diversity of invited speakers and includes gender self-identification options in speaker surveys.} \vspace{.4em}

The Goddard Climate Survey was conducted by the LGBT Advisory Council to survey the workforce of NASA GSFC in 2018 and 2020 (Yargus, K., pers. comm., 2020) and asked participants to self-identify their gender as either male, female, transgender male, transgender female, neither male nor female (genderqueer; nonbinary added in 2020), or self-described (write-in). In 2018 and 2020 respectively, 1.1$\%$ and 0.5$\%$ of respondents identified as neither male nor female.\vspace{.4em}

The Division for Planetary Science (DPS) of the AAS conducted a workforce survey in 2011 [40] but did not ask about nonbinary identities. The survey was updated and repeated in 2019 [41], asking participants to self-identify as either nonbinary or another identity, woman, man, or self-described. 1.1$\%$ of survey respondents identified as nonbinary or another identity.\footnote{Note that here we present bulk answers by student and non-student researchers while in [42] the responses were separated to be comparable with the 2011 DPS survey. See white paper by [43] for further information on LGBTQ+ demographic and workforce issues.}\vspace{.4em}

The Space Science in Context 2020 conference collected demographic data in their attendee feedback survey (Persaud, D.M. and Armstrong, E.S., pers. comm., 2020). Participants were asked to select all options that applied to them: woman, man, nonbinary, trans, and/or self-described. As of July 2020, 13$\%$ of respondents identified as nonbinary. Two thirds of respondents who identified as nonbinary selected at least one additional gender expression. 10$\%$ of presenters and speakers expressed a preference for nonbinary pronouns.\vspace{-.75em}

\subsection{Privacy considerations}

The planetary science community is relatively small and its members are well-connected. Efforts to preserve the privacy and anonymity of survey participants must be taken with extreme care, as demographic data alone can be used to identify individual marginalized members of the community, especially those who face multiple forms of marginalization. For more in-depth discussion of this issue and how to address it in survey design, refer to [3].\vspace{-.75em}

\subsection{Institutional and educational policies and practices}

A full consideration of gender equity in the context of institutional reform is beyond the scope of this paper. Our core recommendation follows that of [3]: \textbf{Gender equity requires the adoption of a more complex model of gender than has historically been employed by equity initiatives. We recommend that planetary science organizations shift their focus from women to people of all marginalized genders.} While this represents a major change to the status quo, as we move into the next decade of planetary science research, it is a change that must be made in order to support all members of the space science community [42, 43, 44].\vspace{.4em}

Planetary scientists need to become familiar with fields such as gender studies, transgender studies, critical race theory, and Science, Technology, and Society Studies (STSS). At the same time, we must recognize the limitations of our knowledge. We suggest that study in these areas be added to undergraduate and graduate curricula and workforce development programs. For additional guidelines on building inclusive planetary science and astrobiology related communities and institutions, we direct readers to the Nashville Recommendations,\footnote{\url{https://tiki.aas.org/tiki-index.php?page=Inclusive_Astronomy_The_Nashville_Recommendations}} the LGBT+ Inclusivity Best Practices Guide [2], and the physics education work of [45].\vspace{-.75em}

\subsection{Final recommendations}


We are no longer asking to be listened to, because we have not been listened to. Instead, \textbf{we are telling you to act}. Do not let suggestions, conversations, or platitudes be the extent of your actions: the only way to bring about change is to not only prioritize the voices and needs of the most marginalized people in our communities, but to take tangible steps toward dismantling unjust systems and materially supporting those most harmed by those systems.\vspace{.4em}

Marginalized people are frequently called upon to educate others about the conditions of their marginalization, typically without compensation [46, 47]. Before consulting marginalized people, we recommend that planetary scientists wishing to educate themselves about gender first seek out resources independently to the best of their abilities, starting with the works cited in this paper.\vspace{-.25em}

\section{Cost estimates}\vspace{-.4em}

The true cost of studies and initiatives like those discussed in this manuscript is an unknown number of students, postdocs, early career researchers, and other scientists and engineers who have been alienated and excluded from the planetary science community.\footnote{We anticipate that any planetary scientist who chooses \textit{not} to conduct such a study will incur costs on the order of \$0.00.}\vspace{.4em}

The hire of social scientists to conduct this work responsibly will, of course, require additional funds. For a brief discussion of the financial costs associated with gender-inclusive survey design, see [48]. We encourage interested planetary science researchers to anticipate and plan for such costs, and to seek out appropriate funding to collaborate with these experts.\vspace{-.9em}

\section*{Acknowledgements}\vspace{-.4em}
We thank the authors of [3] for their guidance during the adaptation of their work and J. Smilges for generous comments that greatly improved this manuscript. The views and opinions expressed in this article are those of the authors and do not necessarily reflect the official policy or position of any agency of the U.S. government. Assumptions made within the analysis are not reflective of the position of any U.S. government entity.\vspace{-.9em}

\section*{References}\vspace{-1.2em}
\begin{multicols}{2}
\justify
\fontsize{10.5}{10.5pt}\fontfamily{ptm}\selectfont
{\textbf{[1]} Atherton et al., American Physical Society (2016). \textbf{[2]} Ackerman et al., ``LGBT+ Inclusivity..." arXiv:1804.08406 (2018). \textbf{[3]} Rasmussen et al., ``The nonbinary fraction..." Astro 2020 Decadal Survey (2019). \textbf{[4]} Berea \& Arcand, ``The role of social sciences..." Astro 2020
Decadal Survey (2019). \textbf{[5]} Rathbun et al., AAS DPS Meeting (2016). \textbf{[6]} Rathbun, Nature Astronomy (2017). \textbf{[7]} Rathbun et al., Planetary Science Vision 2050 Workshop (2017). \textbf{[8]} Rathbun et al., ``What’s new for women+..." (2019). \textbf{[9]} Zellner et al., 51st Lunar and Planetary Science Conference (2020). \textbf{[10]} Pico et al., Earth and Space Science (2020). \textbf{[11]} Clancy et al., Journal of Geophysical Research (Planets) (2017). \textbf{[12]} Richey et al., Bulletin of the AAS (2019). \textbf{[13]} Aycock et al., Physical Review Physics Education Research (2019). \textbf{[14]} Keyes, Proc. ACM Hum.-Comput. Interact. (2018). \textbf{[15]} Grant et al., National Center for Transgender Equality (2011). \textbf{[16]} McLemore, Self and Identity (2015). \textbf{[17]} Davidson, Cogent Social Sciences (2016). \textbf{[18]} Mizock et al., International Journal of Transgenderism (2017). \textbf{[19]} Thorpe, Antistasis (2017). \textbf{[20]} Cech \& Pham, Social Sciences (2017). \textbf{[21]} Cech \& Rothwell, Journal of Engineering Education (2018). \textbf{[22]} Combahee River Collective, The Combahee River Collective statement, Kitchen Table: Women of Color Press (1977/1986). \textbf{[23]} Crenshaw, The University of Chicago Legal Forum (1989). \textbf{[24]} Crenshaw, Stanford Law Review (1991). \textbf{[25]} Prescod-Weinstein, ``Intersectionality as a Blueprint..." (2016). \textbf{[26]} Prescod-Weinstein, Nature Astronomy (2017). \textbf{[27]} Sage, The Sociological Review (2009). \textbf{[28]} Oy\v{e}w\`um\'\iı, University of Minnesota Press, (1997). \textbf{[29]} McLelland et al., Routledge, (2005). \textbf{[30]} Chiang, Palgrave Macmillan US (2012). \textbf{[31]} Kugle \& Kugle, NYU Press (2014). \textbf{[32]} Besnier \& Alexeyeff, University of Hawai'i Press (2014). \textbf{[33]} Bhaduri \& Mukherjee, Transcultural Research – Heidelberg Studies on Asia and Europe in a Global Context, Springer India (2015). \textbf{[34]} Monk et al., Nature Ecology \& Evolution (2019). \textbf{[35]} Gonsalves, Physical Review (2018). \textbf{[36]} Ko et al., AIP Conference Series (2013). \textbf{[37]} Horton \& Holbrook, InAIP Conference Proceedings (2013). \textbf{[38]} Guillen et al., AAS Meeting Abstract (2011). \textbf{[39]} Fernandes et al., EarthArXiv Preprints (2020). \textbf{[40]} White, Planetary Science Survey (2011). \textbf{[41]} Hendrix, Rathbun, The DPS Committee, and the DPS Professional Culture \& Climate Subcommittee, 51st LPSC (2020). \textbf{[42]} Rivera-Valent\'{i}n et al., ``Who is Missing in Planetary Science?..." Planetary Science and Astrobiology Decadal Survey 2023-2032 (2020). \textbf{[43]} Vander Kaaden et al., "Creating Inclusive, Supportive, and Safe Environments in Planetary Science..." Planetary Science and Astrobiology Decadal Survey 2023-2032 (2020). \textbf{[44]} Piatek et al., ``Breaking Down Barriers..." Planetary Science and Astrobiology Decadal Survey 2023-2032 (2020). \textbf{[45]} Traxler et al., Physical Review (2016). \textbf{[46]} Fricker, Oxford University Press (2007.) \textbf{[47]} Berenstain, Ergo: An Open Access Journal of Philosophy (2016). \textbf{[48]} Sell, American Journal of Public Health (2017).}
\end{multicols}
\end{document}